\documentclass[aps,prl,reprint,superscriptaddress, showpacs]{revtex4-1}

\usepackage{graphicx}
\usepackage{dcolumn}
\usepackage{bm}

\begin{document}


\title{Inversion symmetry of Josephson current as test of chiral domain wall motion in Sr$_{2}$RuO$_{4}$}


\author{Kohta Saitoh}
\altaffiliation{current affiliation: Department of Applied Physics, Osaka University, Suita 565-0871, Japan}
\author{Satoshi Kashiwaya}
\author{Hiromi Kashiwaya}
\author{Yasunori Mawatari}

\affiliation{National Institute of Advanced Industrial Science and Technology (AIST), Tsukuba, Ibaraki 305-8568, Japan}

\author{Yasuhiro Asano}
\affiliation{Department of Applied Physics, Hokkaido University, Sapporo 060-8628, Japan}

\author{Yukio Tanaka}
\affiliation{Department of Applied Physics, Nagoya University, Nagoya 464-8603, Japan}

\author{Yoshiteru Maeno}
\affiliation{Department of Physics, Kyoto University, Kyoto 606-8502, Japan}

\date{\today}

\begin{abstract}
Clarifying the chiral domains structure of superconducting Sr$_{2}$RuO$_{4}$ has been a long-standing issue in identifying its peculiar topological superconducting state. We evaluated the critical current $I_{c}$ versus the magnetic field $H$ of Nb/Sr$_{2}$RuO$_{4}$ Josephson junctions, changing the junction dimension in expectation of that the number of domains in the junction is controlled. $I_{c}(H)$ exhibits a recovery from inversion symmetry breaking to invariance when the dimension is reduced to several microns. This inversion invariant behavior indicates the disappearance of domain walls; thus, the size of a single domain is estimated at approximately several microns.
\end{abstract}


\maketitle

Strontium ruthenate (Sr$_{2}$RuO$_{4}$; SRO)~\cite{MaenoNature} has long been studied and is now widely accepted as a spin-triplet superconductor. A number of experiments~\cite{IshidaNature, LukeNature, NelsonScience, MaenoJPSJ, MaenoRMP} have supported the pairing state of SRO as spin-triplet chiral $p$-wave with broken time-reversal symmetry. In the chiral $p$-wave symmetry, the orbital part of the pair potential is represented as $k_{x} \pm ik_{y}$, which means that the phase of the pair potential evolves continuously by clockwise or anticlockwise rotation in the $k_{x}$-$k_{y}$ plane reflecting the finite angular momentum of the Cooper pairs. Thus, SRO is believed to be a typical example of a topological superconductor~\cite{KashiwayaPRL, NakamuraPRB, NakamuraJPSJ, JangScience}. In recent years, topological superconductivity has received considerable attention because Majorana fermions, which can be used for topological quantum computation, are expected to emerge in half-quantum vortex cores or at the edges~\cite{ReadPRB, QiRMP, Alicea, TanakaJPSJ}. The search for Majorana fermions is increasingly accelerated. However, the pairing symmetry of SRO is still controversial because some of the features peculiar to the chiral $p$-wave state, such as spontaneous magnetic fields due to the edge currents and chiral domains, have not been observed yet~\cite{KirtleyPRB, HicksPRB}.

To identify the pairing symmetry of unconventional superconductors, the sensitivity of the Josephson effect to the phase of the pair potential is quite useful. In previous high-temperature-superconductor experiments, the magnetic field responses of the critical current $I_{c}$ in corner-shaped Josephson junctions and SQUIDs, whose superconducting loop contains two interfaces with different orientations, have revealed the pairing symmetry to be the $d$-wave state~\cite{HarlingenRMP}. The same idea can essentially be applied to the determination of the chiral $p$-wave state of SRO. However, prior to the detection of the chiral $p$-wave state using the corner-shaped Josephson junctions, the Josephson effect in SRO was not well understood; $i.e.$, the conventional Fraunhofer diffraction pattern~\cite{Barone} of a single-boundary Josephson junction has not been observed yet. One of the reasons for the unconventional behavior is considered to be the effects of chiral domains and their motion during measurement. Kidwingira \textit{et al.} reported a variety of complicated diffraction patterns in Pb/Cu/SRO junctions~\cite{KidwingiraScience}. The interpretation of these complicated diffraction patterns is that the phase of an SRO crystal in a junction is spatially modulated owing to the existence of the chiral domains~\cite{Bouhon}. They also reported several peculiar features indicating the existence of the chiral domain wall motion, such as a telegraph-like noise and a hysteresis loop in the diffraction patterns, and estimated the size of a single chiral domain at approximately 1~$\mu$m.

\begin{figure}[t]
  \begin{center}
    \includegraphics[keepaspectratio=true,height=60mm]{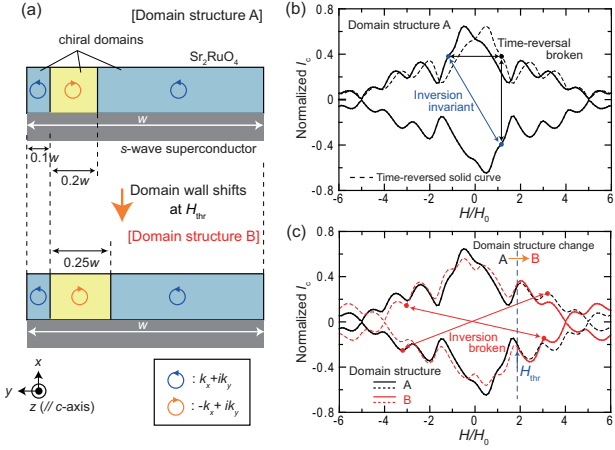}
  \end{center}
  \caption{(a) Chiral domain structures used in the simulation. The two colors represent the different chiral domains ($i.e.$ $\pm k_{x}+ik_{y}$) of SRO. (b) Simulation of $I_{c}(H)$, taking account of the self-field under domain structure A. The critical current $I_{c}$ and the magnetic field $H$ are normalized by $I_{1}$ and $H_{0}$ corresponding to the flux quantum $\Phi_{0}$, respectively. The dashed curve is the time-reversed version of the solid curve. Although the time-reversal symmetry is broken due to the self-field and the chiral domain, the IS is still invariant as long as the domain walls remain static. (c) The IS in $I_{c}(H)$ breaks as the domain wall moves from A to B at the threshold field of $H_{\mathrm{thr}}$ during the field-sweep measurement.}
  \label{Simulation}
\end{figure}

However, the size of a single domain is still a topic under discussion, because the estimated size is largely distributed depending on the experimental probes~\cite{Kallin}. The size larger than 50~$\mu$m was estimated by the polar Kerr effect experiment~\cite{XiaPRL}, while the size as small as $\sim$400~nm was suggested by the scanning SQUID experiment~\cite{HicksPRB}. The determination of the domain size is one of the important issues that could settle the pairing symmetry of SRO. Nelson \textit{et al.} fabricated a AuIn/SRO SQUID in which two junctions were formed at the opposite edges in the $ab$ plane of an SRO crystal and reported the minimum of the magnetic field modulation pattern in $I_{c}$ at zero magnetic field~\cite{NelsonScience}. This result seems to suggest that the pairing symmetry of SRO is odd parity. On the other hand, Asano \textit{et al.} theoretically calculated that the modulation pattern can be shifted in phase by $\pi$, depending on whether the number of domains in the SQUID loop is even or odd~\cite{AsanoSQUID}. Since the size of the SRO crystal they used was on the order of millimeters which probably included a large numbers of domains, the phase shifts at the domain walls should be treated more explicitely.

Here, we report the junction size dependence of the magnetic field $H$ responses of the critical current $I_{c}$ in Nb/SRO Josephson junctions. We expect that the number of domains in the junction is controlled by changing the dimension of the junction, and correspondingly the diffraction pattern $I_{c}(H)$ varies depending on the configuration of domains. We focus on the inversion symmetry (IS) in $I_{c}(H)$, which is invariant in the absence of domain wall motion (the details are given later). As we reduced the width of the junctions, the IS in $I_{c}(H)$ exhibited a recovery from breaking to invariance at a junction width of several microns. This result led us to conclude that the size of a single domain is on the order of several microns.

\begin{figure}[t]
  \begin{center}
    \includegraphics[keepaspectratio=true,height=60mm]{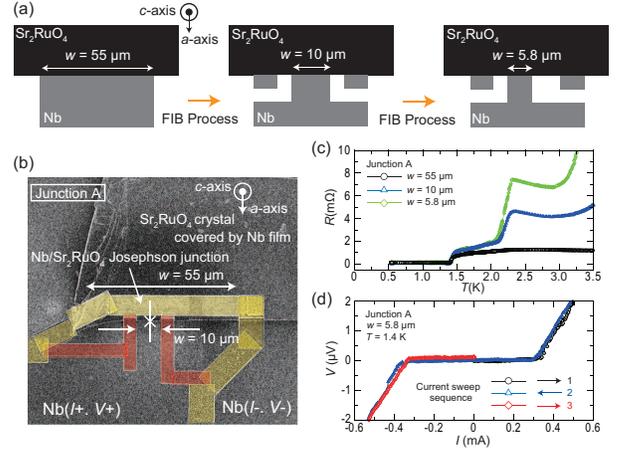}
  \end{center}
  \caption{(a) Schematic of sequence used to modify the junction width $w$. (b) SIM image of junction A at $w$ =10~$\mu$m. Nb films in yellow area were removed to fabricate the initial junction with $w$ = 55~$\mu$m. After measurements at $w$ = 55~$\mu$m, the Nb films in the red areas were removed to change $w$ to 10~$\mu$m. (c) $R$--$T$ characteristics of junction A at each $w$. The superconducting transition temperature $T_{c}$ is approximately 1.4~K at all $w$. (d) Typical $I$--$V$ characteristic of junction A at $w$ = 5.8~$\mu$m observed at $T$ = 1.4~K ($T_{c}\simeq$1.41~K).}
  \label{size}
\end{figure}

First, we illustrate with a simulation the concept behind our experiment. We assume a Josephson junction between a SRO crystal and an $s$-wave superconductor attached at a single side of SRO, and SRO to be the two-dimensional chiral $p$-wave. In the present simulation, we assume that the junction width is sufficiently smaller than the the Josephson penetration depth $\lambda_{J}$ for simplicity. 
As shown in Fig.~\ref{Simulation}(a), we employ the chiral-domain model in which the $y$-component (parallel to the interface) of the pair potential keeps its phase, while the $x$-component (perpendicular to the interface) changes its phase by $\pi$ at the domain boundary~\cite{AsanoSQUID}.  $I_{c}(H)$ is evaluated by taking the Josephson current $I$ as the form of $I=I_{1}\cos \theta -I_{2}\sin 2\theta$, where $I_{1} \gg I_{2}$ and $\theta$ is the phase of SRO relative to that of the $s$-wave~\cite{AsanoJJ, AsanoSQUID}. We calculate $I_{c}(H)$'s for positive and negative current directions represented by $I_{c}^{+}(H)$ and $I_{c}^{-}(H)$, respectively, in order to evaluate the symmetry of $I_{c}(H)$ with respect to the field and the current direction. When both the chiral domains and the self-field are absent, $I_{c}(H)$ is time-reversal invariant $i.e.$, $I_{c}^{\pm}(H)=I_{c}^{\pm}(-H)$ and $I_{c}^{+}(H)=-I_{c}^{-}(H)$. In contrast, when the effect of the self-field cannot be neglected, the $I_{c}(H)$ calculated for chiral domain structure A in Fig.~\ref{Simulation}(a) exhibits broken time-reversal symmetry, as shown in Fig.~\ref{Simulation}(b), while the $I_{c}(H)$ is still IS-invariant, $i.e.$, $I_{c}^{+}(H)=-I_{c}^{-}(-H)$, as far as the chiral domains remain static. On the other hand, if once the chiral domain wall shifts from A to B as increasing $H$ beyond a threshold field $H_{\mathrm{thr}}$ at which the domain walls begin to move (Fig.~\ref{Simulation}(a)), the $I_{c}(H)$ is modified from the black curves to the red curves as shown in Fig.~\ref{Simulation}(c), and then $I_{c}(H)$ is no longer invariant with respect to the IS. Therefore, the chiral domain motion can be detected sensitively by testing the IS invariance in $I_{c}(H)$. We note that the IS is insensitive to the junction dimension, $\lambda_{J}$, and the uniformity of the current. According to the previous experiments, chiral domain wall motion is considered to be excited by applied magnetic fields~\cite{KidwingiraScience} or by a current flow~\cite{KambaraPRL, KambaraJPSJ, Anwar}. Here, we aim to test the domain wall motion by reducing the junction dimension.

\begin{figure}[t]
  \begin{center}
   \includegraphics[keepaspectratio=true,width=85mm]{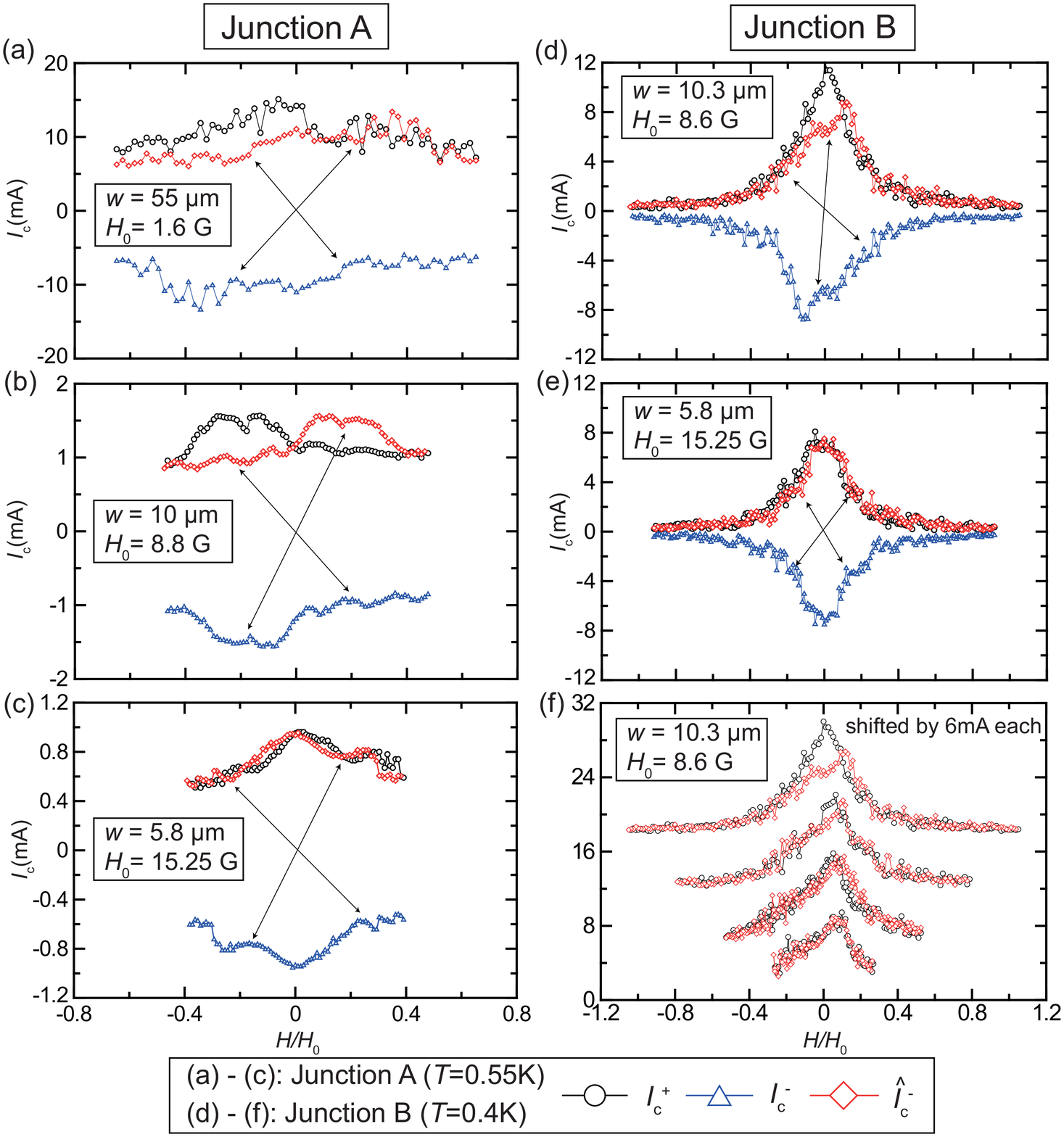}
  \end{center}
  \caption{Magnetic field $H$ responses of critical current $I_{c}(H)$ in junction A ($T$=0.55~K) at (a) $w$ = 55~$\mu$m, (b) $w$ = 10~$\mu$m and (c) $w$ = 5.8~$\mu$m, and in junction B ($T$=0.4~K) at (d)$w$=10.3~$\mu$m and (e)$w$=5.8~$\mu$m. The applied magnetic field $H$ (//$c$-axis) is normalized by the period for the conventional Fraunhofer pattern of $H_{0}$ estimated for each $w$ shown in the figures. $I_{c}^{+}(H)$ (black data) and $I_{c}^{-}(H)$ (blue data) are $I_{c}(H)$'s in a positive and a negative current directions, respectively. $\hat{I}_{c}^{-}(H)$ (red data) is obtained from $I_{c}^{-}(H)$ under the inverse projection with respect to the current direction and $H$ (represented as black arrows). The IS of the junctions is evaluated by the consistency between $I_{c}^{+}(H)$ and $\hat{I}_{c}^{-}(H)$. In contrast to the results at (a), (b) and (d), $\hat{I}_{c}^{-}(H)$ is consistent with $I_{c}^{+}(H)$ at $w$ = 5.8~$\mu$m in both junction A and B [(c) and (e)]. (f)$H$-sweep-range dependence of the IS in junction B at $w$=10.3~$\mu$m. The IS gradually recovered by reducing the sweep range, and $H_{\mathrm{thr}}$ is estimated at $H/H_{0}\sim0.25$(=2.1~G).}
  \label{Size-dep}
\end{figure}

\begin{figure}[t]
  \begin{center}
    \includegraphics[keepaspectratio=true,width=85mm]{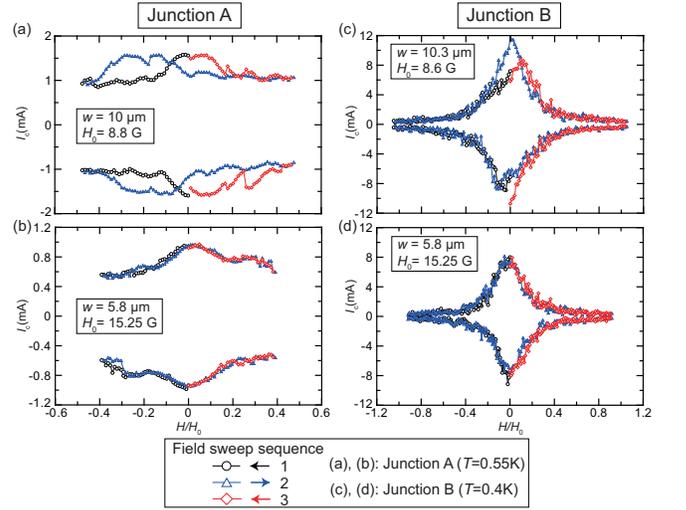}
  \end{center}
  \caption{Hysteresis loops in $I_{c}(H)$ characteristics. Magnetic field $H$ was swept from zero to a negative value (black data), and then swept up to a positive value (blue data), and finally swept back to zero (red data). Diffraction patterns observed at $w \sim$ 10~$\mu$m showed hysteresis loops depending on the direction in which the magnetic field was swept [(a) and (c)]. However, the hysteresis loops disappeared in both junction A and B as we reduced $w$ to 5.8~$\mu$m [(b) and (d)].}
  \label{Hysteresis}
\end{figure}

Next, we move to the experimental side. We succeeded in fabricating Nb/SRO Josephson junctions in which high supercurrent densities (as high as on the order of 10$^{7}$~A/m$^{2}$) are realized~\cite{SaitohAPEX}. In the present work, the width of junctions was sequentially modified using a focused ion beam (FIB) so that we can clarify purely the effect of the junction dimension. Figure \ref{size}(a) shows the schematic illustration of the typical sequence used to modify the width of the junction. We fabricated junctions with widths $w$ successively made narrower from $w$ = 55~$\mu$m to 10~$\mu$m, and to 5.8~$\mu$m (junction A), and from $w$=10.3~$\mu$m to 5.8~$\mu$m (junction B). A scanning ion microscopy (SIM) image of junction A at $w$ = 10~$\mu$m is shown in Fig.~\ref{size}(b). After we measured the transport properties of the junctions at the wider size, the junction width $w$ was successively changed to the narrower size. The transport properties of the junctions at each $w$ were measured using a standard four-terminal method down to approximately 0.4~K. The junctions were magnetically shielded by using double $\mu$-metal shields to reduce the residual magnetic field less than 4~mG. The maximum critical current $I_{c}$ shown later in Table~\ref{table} is almost proportional to $w$. Thus, our junctions are considered to be mostly uniform. The $\lambda_{J}$'s estimated by the critical current densities are approximately 5~$\mu$m (junctin A)~\cite{SaitohAPEX} and 3~$\mu$m (junction B), respectively. Figure \ref{size}(c) shows the resistance--temperature ($R$--$T$) characteristics of junction A at each $w$. Although the resistance at a normal state increased as $w$ decreased, a sharp superconducting transition was maintained at $T_{c}\sim$ 1.4~K for all $w$. This result confirms that the FIB process for modifying $w$ did not damage the junction quality. As the reduction of the resistance at $T\sim$2.3~K is clearly separated from the transition at $T_{c}\sim$1.4~K, we consider that the 3-K phase~\cite{MaenoRMP, MaenoJPSJ} in the bulk SRO crystal near the junction reduced the resistance at $T\sim$2.3~K. Figure \ref{size}(d) shows a current--voltage ($I$--$V$) characteristic of junction A at $w$ = 5.8~$\mu$m observed at $T$=1.4~K. The $I$--$V$ characteristic exhibited a typical overdamped behavior with no hysteresis loop.

Figure \ref{Size-dep} shows the magnetic field $H$ responses of the critical current $I_{c}$ in junction A and B. The applied  field (//$c$-axis) swept up from a negative to a positive value is normalized by $H_{0}$, which is the period for the conventional Fraunhofer pattern estimated for each $w$; $H_{0}$ = $\Phi_{0}/\left[w(\lambda_{\mathrm{SRO}}+\lambda_{\mathrm{Nb}})\right]$, where $\Phi_{0}$ is the flux quantum (20.7~G$\cdot \mu$m$^{2}$), and $\lambda_{\mathrm{SRO}}$ (= 190~nm for $H$//$c$-axis) and $\lambda_{\mathrm{Nb}}$ (= 44~nm) are the penetration depths in SRO and Nb, respectively~\cite{SaitohAPEX}. 
The consistency or inconsistency between $I_{c}^{+}(H)$ and $\hat{I}_{c}^{-}(H)$ determines the IS invariance or breaking, respectively, where $\hat{I}_{c}^{-}(H)$ was obtained from $I_{c}^{-}(H)$ by the inverse projection, $i.e.$, $\hat{I}_{c}^{-}(H) = -I_{c}^{-}(-H)$.
In junction A, $I_{c}(H)$ at $w$ = 55~$\mu$m tended to change irregularly, and we have not observed any periodic $I_{c}(H)$ [Fig.~\ref{Size-dep}(a)]. This behavior is reasonable because many chiral domains are considered to be present inside the junction area, and simultanously the width of $w$ = 55$\mu$m is much larger than $\lambda_{J}\sim 5~\mu$m.
As the width of the junction was reduced to $w$ = 10~$\mu$m, the $I_{c}(H)$ tended to exhibit a peak structure, although the maximum of $I_{c}^{+}(H)$ and the minimum of $I_{c}^{-}(H)$ shifted to a negative $H$ as shown in Fig.~\ref{Size-dep}(b). The reduction of the peak width from the expected value ($H/H_{0}\sim$0.15) is attributed to the concentration of the applied magnetic field at the edge of the SRO crystal due to the Meissner effect~\cite{SaitohAPEX}. As further reducing the junction width to $w$ = 5.8~$\mu$m, the $I_{c}(H)$ became rather conventinal as shown in Fig.~\ref{Size-dep}(c); the maximum of $I_{c}^{+}(H)$ and the minimum of $I_{c}^{-}(H)$ were observed at $H/H_{0}\sim0$. Moreover, $\hat{I}_{c}^{-}(H)$ became almost consistent with $I_{c}^{+}(H)$, indicating the recovery of the IS invariance. This feature is quite different from those observed on $w$ = 55~$\mu$m and 10~$\mu$m junctions. Similar recovery of the IS has also been observed in junction B, $i.e.$, the IS was broken at $w$ = 10.3~$\mu$m due to the difference between $I_{c}^{+}(H)$ and $\hat{I}_{c}^{-}(H)$ around $H/H_{0}\sim0$ [Fig.~\ref{Size-dep}(d)]. However, a conventional pattern with the IS invariance was recovered at $w$ = 5.8~$\mu$m, as shown in Fig.~\ref{Size-dep}(e). In order to confirm the validity of the chiral-domain model, we further evaluate the threshold field $H_{\mathrm{thr}}$ in junction B. Figure ~\ref{Size-dep}(f) shows the magnetic field-sweep-range dependence of $I_{c}(H)$. The data of the largest sweep range in Fig.~\ref{Size-dep}(f) is same as that shown in Fig.~\ref{Size-dep}(d). The gradual recovery of the IS invariance by reducing the field-sweep range means the suppression of chiral domain motion in smaller field. The complete recovery of the IS in the lowest curve indicates that the $H_{\mathrm{thr}}/H_{0}\sim0.25$ ($H_{\mathrm{thr}}$=2.1~G) in junction B. Therefore, the recovery of the IS invariance at $w$ = 5.8~$\mu$m under the field-sweep range of $\pm14$~G, which is far larger than $H_{\mathrm{thr}}$, shown in Fig.~\ref{Size-dep}(e) suggests the absence of domain walls inside the junction area.

In addition to the IS, we detected the disappearance of a hysteresis loop in both junction A and B as $w$ was reduced. The magnetic field was swept from zero to a negative value, and then swept up to a positive value, and finally swept back to zero. At $w \sim$ 10~$\mu$m, we observed a hysteresis loop depending on the direction in which the magnetic field was swept [Fig.~\ref{Hysteresis}(a) and (c)]. In Fig.~\ref{Hysteresis}(a), the maximum of $I_{c}^{+}(H)$ and the minimum of $I_{c}^{-}(H)$ shift in the direction in which the magnetic field was swept. In Fig.~\ref{Hysteresis}(c), the magnitude of the maximum $I_{c}^{+}(H)$ and the minimum $I_{c}^{-}(H)$ changes depending on the sweep direction. These hysteresis loops disappeared at $w$ = 5.8~$\mu$m in the both junction A and B, as shown in Fig. \ref{Hysteresis}(b) and (d). Accepting that the origin of the hysteresis loop is the chiral domain wall motion\cite{KidwingiraScience, Bouhon}, the chiral domains are considered to be present at $w \sim$ 10~$\mu$m, while they disappear at the junctions of 5.8~$\mu$m. 

\begin{table}[t]
 \caption{Summary of junction size dependence of magnetic field $H$ responses of critical current $I_{c}$. The value of maximum $I_{c}$ is averaged over several cooling cycles, and $\Delta I_{c} / I_{c}$, where $\Delta I_{c}$ is the standard deviation, is estimated.} 
 \begin{center}
  \begin{tabular}{|c|c|c|c|c|c|}
    \hline
     Junction & $w$ [$\mu$m]  & $I_{c}$ [mA] & $\Delta I_{c} /I_{c}$ [\%] & $I_{c}(H)$   &  IS  \\
    \hline 
      A & 55 &  9.33 & 8.39  &  Random  &  $\times$   \\
    \cline{2-6}
      ($T$=0.55~K) & 10 &  1.55 & 1.09  & Hysteresis   & $\times$   \\
    \cline{2-6}
       & 5.8&  0.94 & 0.58  &  Conventional  &  $\bigcirc$   \\
    \hline
      B & 10.3 & 11.71 & 2.64 & Hysteresis & $\times$ \\
    \cline{2-6}
      ($T$=0.4~K) & 5.8 & 7.96 & 1.33 & Conventional & $\bigcirc$ \\
    \hline
  \end{tabular}
 \end{center}
 \label{table}
\end{table}
 
Table~\ref{table} summarizes the results of the junction size dependence of $I_{c}$. In addition to the above mentioned features, we discuss the distribution of $I_{c}$ estimated by $\Delta I_{c}/ I_{c}$, where $\Delta I_{c}$ is the standard deviation of $I_{c}$ over several cooling cycles. Assuming that chiral domain textures are expected to be inequivalent in each cooling cycle, the variation in $I_{c}$ reflects the presence of multiple numbers of chiral domains and the variation of their configuration. Thus, the tendency that $\Delta I_{c} / I_{c}$ decreases as $w$ is reduced reflects that the number of chiral domains decreases as $w$ is reduced. Putting together the IS invariance and the lack of the hysteresis loop in both junction A and B as $w$ is reduced, we conclude that the detected size dependence of $I_{c}(H)$ is governed  by the chiral domains and their motion, and that the size of a single chiral domain is estimated on the order of several microns. 

The domain size is almost consistent with those estimated by several results using the 3-K phase~\cite{KambaraPRL, KambaraJPSJ, Anwar}, while it is somewhat larger than $\sim$1~$\mu$m estimated by Kidwingira \textit{et al.}~\cite{KidwingiraScience}. We speculate that the relatively larger domain can be induced by $s$-wave Nb films whose superconducting transition temperature $T_{c}$ ($\sim$6.5~K) is higher than that of SRO. In our Josephson junctions, the contact between the Nb films and SRO realizes a high current density~\cite{SaitohAPEX} compared to that of other junctions using the 1.5-K phase. Thus, the phase of SRO was locked to that of the Nb film through the Josephson coupling, which probably results in the creation of the relatively large size of domains. The effect of the phase lock can be checked by using $s$-wave superconductors whose $T_{c}$ is lower than that of SRO, such as Al (typical $T_{c}\sim$1.2~K). This phase-lock effect also might be associated with the difference in $I_{c}(H)$ over multiple cooling cycles. The $I_{c}(H)$ of junction B ($w$=5.8~$\mu$m) was completely stable against the several cooling cycles, whereas that of junction A ($w$=5.8~$\mu$m) showed a different pattern for the rare occasion. Since the current density of junction B is about one order of magnitude higher than that of junction A, it is reasonable to conclude that the stable $I_{c}(H)$ in junction B is also attributed to the phase-lock effect.

Although the chiral $p$-wave symmetry of SRO has been assumed throughout this paper, the present result does not exclude the possibility of the helical $p$-wave symmetry that is another candidate of the paring symmetry of SRO~\cite{MaenoJPSJ}. We believe that the analyses are mostly unchanged if we consider the presence of helical domains instead of chiral domains. For the analysis based on the helical domain models, theoretical calculation performed on the helical domain boundaries is required.

In summary, we measured the junction size dependence of the magnetic field $H$ responses of the critical current $I_{c}$ in Nb/Sr$_{2}$RuO$_{4}$ Josephson junctions and tested the inversion symmetry (IS) invariance of $I_{c}(H)$. The IS exhibited a recovery from breaking to invariance at a junction width of several microns. This inversion invariant recovery indicates the absence of chiral domain wall motion and led us to conclude that the size of a single chiral domain is on the order of several microns. These results will open the possibility that the internal phase of Sr$_{2}$RuO$_{4}$ can be identified by using corner-shaped junctions with a size smaller than several microns in the future.

\begin{acknowledgments}
We are grateful to M.~Koyanagi for the fabrication of the junctions. We thank S.~Yonezawa, and M.~S.~Anwar for fruitful discussions. This work was supported by MEXT KAKENHI Grant Numbers 22103002,15H05852 and 15H05853, and by Grants-in-Aid for Scientific Research (No. 60443181 and No. 15H03686) from the Japan Society for the Promotion of Science, Japan.

\end{acknowledgments}


%

\end{document}